\documentclass[12pt,onecolumn]{article}

\usepackage[utf8]{inputenc} 
\usepackage[T1]{fontenc}    
\usepackage{hyperref}       
\usepackage{url}            
\usepackage{booktabs}       
\usepackage{amsfonts}       
\usepackage{nicefrac}       
\usepackage{microtype}      
\usepackage{lipsum}         
\usepackage{graphicx}       
\usepackage{amsmath}        
\usepackage{setspace}       
\usepackage{natbib}
\usepackage{multirow}

\usepackage{geometry}
\newcommand{\kruskal}[1]{\left[\!\left[#1\right]\!\right]}
\newcommand{\bigsum}{\mathop{\vcenter{\hbox{\scalebox{2.0}{$\displaystyle\sum$}}}}}

\title{Efficient Spatial-Spectral Feature Extraction in Hyperspectral Images via Holistic Multivariance Decomposition}

\author{
 \begin{tabular}{c}
    Süha Tuna\thanks{Corresponding author.} \\
    \small Informatics Institute, Department of Computational Science and Engineering,  \\
    \small \.Istanbul Technical University, \.Istanbul 34469, T\"urkiye \\
    \small \texttt{suhatuna@itu.edu.tr}
  \end{tabular}
}

\date{}

\begin{document}

\maketitle

\begin{abstract}
Tensor decomposition serves as a foundational tool for feature extraction in hyperspectral image classification, a domain classically dominated by the Tucker and Canonical Polyadic decompositions. Although widely adopted, these schemes often struggle to fully encapsulate the deeply coupled geometric and intrinsic  structures inherent to multidimensional hyperspectral data. Their structural reliance on rigid low-rank approximations successfully captures independent mode variations but systematically neglects complex, cross-domain spatial and spectral interdependencies. To overcome this limitation, we introduce the Holistic Multivariance Decomposition framework to achieve highly discriminative hyperspectral feature extraction. The Holistic Multivariance Decomposition provides a novel, structurally flexible tensor algorithm that explicitly models isolated spatial and spectral behaviors, alongside intricate, higher dimensional cooperative interactions. Comprehensive experimental evaluations across four benchmark hyperspectral datasets demonstrate that the proposed Holistic Multivariance Decomposition approximants consistently yield superior classification accuracy compared to conventional Tucker and Canonical Polyadic decomposition methods across diverse supervised learning algorithms. By effectively preserving essential joint multivariance features even under severe subspace compression, these results establish the Holistic Multivariance Decomposition as a robust, high fidelity computational framework for resolving complex multidimensional data structures in hyperspectral analysis.
\end{abstract}


\section{Introduction}
\label{sec:intro}

Hyperspectral (HS) imaging represents a rapidly advancing frontier in remote sensing and data acquisition. These complex data structures are typically captured via unmanned aerial vehicles (UAVs), drones, and satellites operating at various altitudes. By integrating rich informational content across both the spatial and spectral domains, hyperspectral imagery (HSI) embeds comprehensive information specific to a target geographic area \cite{hsi_overview}.

An HS image consists of hundreds of narrow, contiguous spectral bands. These bands facilitate the precise differentiation of distinct objects within an image frame. This high level of discrimination is possible because every material possesses a unique spectral signature. Because these detailed spectral signals provide fundamental information regarding underlying material components, they are widely exploited to resolve material classification problems. Consequently, the task of HS image classification has emerged as a crucial area of inquiry that has consistently attracted significant scientific attention over the past decade \cite{hs_apps_review}.

Beyond traditional remote sensing, HSI has found a wide variety of applications across diverse scientific disciplines. In agricultural and various industrial sectors, notable implementations of HSI include the assessment of food quality and safety \cite{hs_food}, as well as artwork authentication \cite{artwork} and the examination of counterfeit pharmaceuticals \cite{drug}. The technology is also gaining traction in biomedical engineering fields. Here, HSI has been successfully employed for the classification of corneal epithelium injuries \cite{corneal} and the diagnosis of gastric cancer \cite{gastric}. Whether applied to environmental monitoring or clinical diagnostics, the dual presence of spatial and spectral data ensures robust pattern recognition \cite{pattern}, detailed image classification \cite{hs_class}, and precise spectral unmixing \cite{unmixing}.

To improve overall classification accuracy, a diverse array of feature extraction techniques has been developed and implemented. Broadly speaking, these methodologies fall into two primary groups, conventional mathematical schemes and advanced learning based approaches. Within conventional frameworks, many data processing techniques originally designed for standard signals and images have been adapted for HS imagery.

For instance, classical data reduction techniques like principal component analysis (PCA) \cite{pca}, linear discriminant analysis (LDA) \cite{lda}, and factor analysis (FA) \cite{factor} are frequently applied along the spectral axis of HS images. These methods extract localized features from the corresponding spectral signals quite effectively. Although these techniques function well as feature extractors and yield satisfactory results, they fundamentally struggle to capture the complex spatial correlations inherent to HSI data.

To address this spatial limitation, researchers often apply standard feature extraction techniques designed for conventional color images. These methods are run sequentially on each individual spectral band to elicit structural information from the HS cube. Within this sequential paradigm, transform based approaches such as the discrete wavelet transform (DWT) \cite{tuna_wavelet_hdmr} and the discrete cosine transform (DCT) \cite{dct} serve as vital analytical tools. However, despite their acknowledged success in localized feature extraction, these methods possess an inherent two dimensional nature. Consequently, they fail to simultaneously capture critical inter-band correlations across the spectral domain.

An alternative paradigm for feature extraction in HSI relies on tensor decomposition based techniques. To this end, well known models such as the Tucker \cite{tucker_hsi,kolda,sidiropoulos} and Canonical Polyadic (CP) \cite{cp_hsi,kolda,sidiropoulos} decompositions have been widely applied to HSI data. However, because these conventional methods depend on rigid low-rank approximations, they often fail to adequately capture complex mode interrelations among distinct dimensions, particularly joint spatial–spectral features. To address this limitation, High Dimensional Model Representation (HDMR) \cite{tuna_wavelet_hdmr,tuna_sparse_hdmr,efe_fusion} and Enhanced Multivariance Products Representation (EMPR) \cite{tuna_tgrs,enis} frameworks have been effectively employed to isolate these spatial–spectral correlations. Unfortunately, both HDMR and EMPR lack flexible rank adjustment mechanisms, inherently restricting the scope of the extracted information. To overcome these bottlenecks, we propose a novel framework named Holistic Multivariance Decomposition (HMD). The proposed HMD not only extracts highly efficient spatial–spectral features from HS images but also introduces a controllable dimensionality reduction parameter, allowing for dynamic adjustment of the retained subspace information.

The remainder of this manuscript is organized as follows. Section \ref{sec:bg} briefly reviews the fundamentals of tensor decompositions and their applications in hyperspectral image (HSI) processing. Section \ref{sec:hmd} introduces the proposed framework, detailing the mathematical formulations of the Holistic Multivariance Decomposition (HMD) method. The experimental setup, benchmark datasets, and comparative results are presented in Section \ref{sec:results} Finally, Section \ref{sec:conc} concludes the paper with a summary of key findings and future research directions.

\section{Background}
\label{sec:bg}

In remote sensing and earth observation, HSI yields massive three dimensional data cubes. These structures are characterized by highly coupled spatial and spectral modes. Preserving these inherent geometric structures during feature extraction requires advanced multilinear algebra frameworks such as tensor decomposition \cite{kolda,sidiropoulos}. Crucially, these frameworks must remain capable of maintaining vital spatial-spectral correlations. Consequently, researchers rely on computationally efficient tensor decompositions to uncover latent patterns. The mathematical foundations of this domain rest primarily on two approaches. The first is the Tucker decomposition. This model isolates the principal subspaces of the data by decomposing the tensor into a compact core tensor multiplied by orthogonal factor matrices along each independent dimension \cite{tucker_hsi}. The second is the Canonical Polyadic (CP) decomposition \cite{cp_hsi}, which represents a higher order tensor as a minimal sum of rank-one vector outer products. 

A substantial portion of modern feature extraction relies on low-rank approximation phenomena. These techniques aim to reconstruct high dimensional tensors using a truncated summation of elementary components \cite{kolda,sidiropoulos}. In classical Tucker and CP frameworks, these components are structurally established by computing the outer product of individual vectors drawn from each independent dimension (mode). While this classical paradigm is mathematically elegant and computationally tractable, it suffers from a severe structural limitation. Since the underlying terms are rigidly tied to pure multilinear outer products, each rank-one component can only capture the isolated, independent variations of individual modes. Therefore, traditional low-rank approximations inherently overlook the complex, mutual relationships and highly coupled interdependencies that characterize the nature of HS data. When applied to HS datasets, this independence assumption forces the decomposition to sacrifice structural fidelity. Ultimately, this limitation prevents the framework from capturing the shared cross-mode dynamics, particularly the essential spatial-spectral correlations necessary for robust feature extraction \cite{tensor_review}.

To overcome the limitations associated with assuming strictly decoupled modal (e.g., spatial–spatial or spatial–spectral) behaviors, recent research has pivoted toward multidimensional analysis frameworks explicitly engineered to map higher order interactions. Originating in function approximation, techniques such as HDMR decompose complex systems by systematically separating independent, single mode influences from their cooperative, higher dimensional interactions \cite{tuna_wavelet_hdmr, efe_fusion, tuna_sparse_hdmr}. To enhance the representation capability of HDMR, the EMPR framework was introduced to provide a more descriptive representation of tensors. Unlike conventional tensor decompositions that rely strictly on rigid basis matrices, EMPR integrates flexible, one dimensional preselected support vectors directly into its structural representation \cite{tuna_tgrs, enis}. 

Regardless of whether preselected or data-driven optimized support vectors are employed, the representation capability of EMPR remains inherently constrained \cite{tuna_tgrs, enis}. Because these constituent support structures are strictly one dimensional, they can only capture information along a single direction at each mode. This structural limitation introduces a critical bottleneck in EMPR implementations. To overcome this restriction and enhance the framework's capacity to preserve both spatial–spatial and spatial–spectral interactions within a given HS image, it is essential to transition from vector valued to matrix valued supports. To this end, we propose a novel framework termed Holistic Multivariance Decomposition (HMD).

\section{Holistic Multivariance Decomposition for HSI}
\label{sec:hmd}

HS images are naturally represented as a third order tensor $\mathcal{H} \in \mathbb{R}^{n_1 \times n_2 \times n_3}$, where the first two modes denote spatial dimensions and the third mode represents the spectral bands. To adapt the Holistic Multivariance Decomposition (HMD) framework for HS data, we define a set of mode specific support matrices $\mathbf{U}_i$, subject to the dimensionality reduction parameters $r_i$ such that
\begin{equation}
\mathbf{U}_i \in \mathbb{R}^{n_i \times r_i}\,; \qquad r_i < n_i\,, \qquad i=1, 2, 3
\label{eq:Si}
\end{equation}
These support matrices are constrained to be scaled semi-orthogonal, satisfying the following condition
\begin{equation}
\label{eq:Sorth}
\mathbf{U}_i^T\, \mathbf{U}_i = n_i\,I_{r_i \times r_i}; \qquad i=1, 2, 3
\end{equation}
To simplify the subsequent derivations, let $\mathcal{U}$ denote the complete set of support matrices as follows
\begin{equation}
\mathcal{U} = \left\{\mathbf{U}_1, \mathbf{U}_2, \mathbf{U}_3\right\}
\end{equation}
We define $\mathcal{U}^{(i)}$ and $\mathcal{U}^{(i,j)}$ as the specific subsets that exclude the projection matrices corresponding to the $i$-th mode, or both the $i$-th and $j$-th modes as 
\begin{equation}
\label{eq:Sis}
\mathcal{U}^{(i)} = \left\{\mathbf{U}_k\, \vert \, k \neq i \right\}; \qquad i=1, 2, 3
\end{equation}
and 
\begin{equation}
\label{eq:Sijs}
\mathcal{U}^{(i,j)} = \left\{\mathbf{U}_k \, \vert \, k \neq i, j \right\} ; \qquad i=1, 2, 3; \qquad i < j
\end{equation}
respectively. 

Utilizing the Tucker operator \cite{kolda}, the $3$-D HMD expansion for the HS tensor $\mathcal{H}$ decomposes the data into a hierarchical structure of multivariance components. This isolates baseline interactions, single-mode directional variations, and complex spatial-spectral interrelations
\begin{equation}
\mathcal{H} = \kruskal{\mathcal{H}_0\,;\, \mathcal{U} } + \displaystyle\sum_{i=1}^3 \kruskal{ \mathcal{H}_i\,;\, \mathcal{U}^{(i)} }
  + \displaystyle\sum_{\substack{i,j=1 \\ i < j}}^3 \kruskal{ \mathcal{H}_{i,j}\,;\, \mathcal{U}^{(i,j)} } 
 + \mathcal{H}_{1,2,3} 
\label{eq:hmd}
\end{equation}
where the explicit definition of the Tucker operator $\kruskal{\,\cdot\,}$ is given as follows
\begin{equation}
\label{eq:tucker}
\kruskal{\mathcal{H}\,;\, \mathbf{A}, \mathbf{B}, \mathbf{C}}
=\mathcal{H}\times_1\mathbf{A}\times_2\mathbf{B}\times_3\mathbf{C}
\end{equation}
as long as the dimensions of the tensor and factor matrices hold.

In (\ref{eq:hmd}), $\mathcal{H}_0$ is called the zeroth degree HMD component, the $\mathcal{H}_i$ terms are named the first degree HMD components, while the $\mathcal{H}_{i,j}$ terms are called the second-degree HMD components. The last term in (\ref{eq:hmd}) is named the residual term. The relevant HMD terms are achieved by multiplying each HMD component with the corresponding support matrices across the corresponding modes. By truncating the HMD expansion in (\ref{eq:hmd}) at a certain level, the corresponding HMD approximants are established. If only the zeroth HMD term is taken into account, then the zeroth level HMD approximant is obtained. Likewise, if the first order HMD terms are added to the zeroth level HMD term, then the first level HMD approximant is established. Finally, if the second order HMD terms are also taken into consideration alongside the previous HMD terms, the second level HMD approximant is achieved.   
 
\begin{figure}[!h]
\centering
\includegraphics[scale=.22]{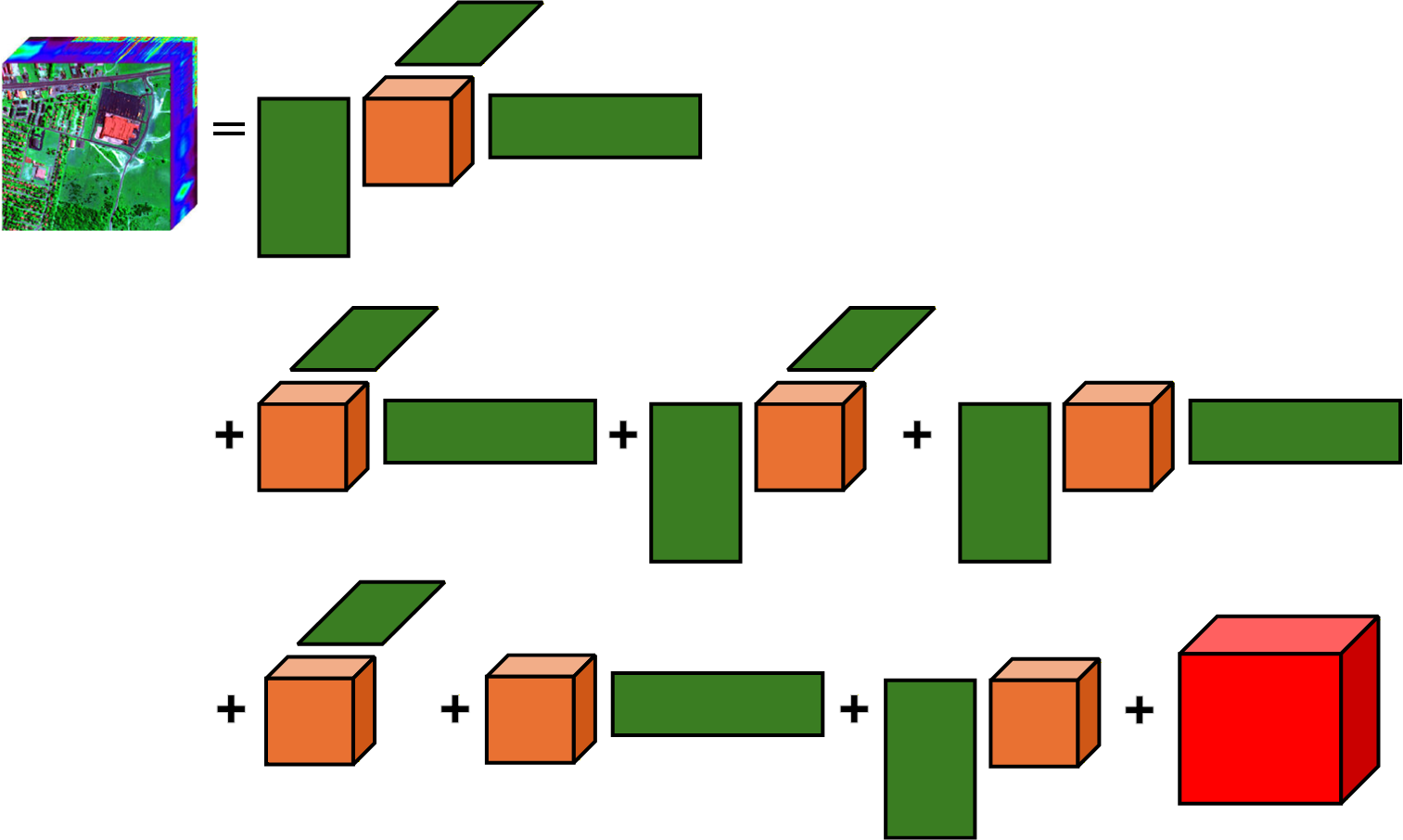}
\caption{Visual representation of the hierarchical $3$-D HMD expansion for hyperspectral data cubes, illustrating the isolation of zeroth, first, and second degree multivariance components.}
\label{fig:hmd3d}
\end{figure}

A geometric illustration of this $3$-D HMD expansion is displayed in Fig. \ref{fig:hmd3d}. Unlike HDMR aand EMPR techniques, HMD ensures that each isolated component remains a full three dimensional tensor for $3$-D case. To systematically compute these individual components, we exploit the fundamental properties of the tensor mode-$n$ product \cite{kolda}, specifically its commutativity and identity relations with respect to the support matrices such as

\begin{equation}
\label{eq:modenm}
\mathcal{H}\times_m \mathbf{A}\times_n \mathbf{B} = \mathcal{H}\times_n \mathbf{B}\times_m
\mathbf{A},\qquad m\neq n
\end{equation}
and
\begin{equation}
\label{eq:modenn}
\mathcal{H}\times_n \mathbf{U}_i\times_n \mathbf{U}_i = \mathcal{H}\times_n \left(\mathbf{U}_i^T\,\mathbf{U}_i\right),\qquad i=1,2,3
\end{equation}
respectively.

By applying the Tucker operator using the complete support set $\mathcal{U}$ to both sides of the expansion and applying the properties given in (\ref{eq:modenm}) and (\ref{eq:modenn}), the zeroth level HMD term $\mathcal{H}_0$ can be isolated. Geometrically, rather than collapsing the data, this term safely compresses the HS cube into a dense spatial-spectral core tensor that preserves the global multidimensional nature as follows
\begin{equation}
\label{eq:X0}
\mathcal{H}_0 = \frac{1}{n_1n_2n_3} \kruskal{ \mathcal{H}\,;\, \mathbf{U}_1,\mathbf{U}_2,\mathbf{U}_3 }
\end{equation}
where $\mathcal{H}_0$ is a tensor of size $r_1\times r_2\times r_3$. 

Following a similar deflation strategy, the first level HMD terms are obtained via partial projection. These terms isolate distinct, linear variations along a single axis such as pure spectral deviations ($\mathcal{H}_3$) or localized spatial edge structures ($\mathcal{H}_1, \mathcal{H}_2$). They behave geometrically as bundles of parallel fibers within the tensor subspace. Therefore, the explicit definitions of the first degree HMD terms are given as follows
\begin{eqnarray}
\label{eq:Xi}
\mathcal{H}_1 &=& \frac{1}{n_2n_3} \kruskal{ \mathcal{H}\,;\, \mathbf{U}_2, \mathbf{U}_3} - 
\kruskal{\mathcal{H}_0\,;\, \mathbf{U}_1} \nonumber \\
\mathcal{H}_2 &=& \frac{1}{n_1n_3} \kruskal{ \mathcal{H}\,;\, \mathbf{U}_1, \mathbf{U}_3} - 
\kruskal{\mathcal{H}_0\,;\, \mathbf{U}_2} 
\nonumber \\
\mathcal{H}_3 &=& \frac{1}{n_1n_2} \kruskal{ \mathcal{H}\,;\, \mathbf{U}_1, \mathbf{U}_2} - 
\kruskal{\mathcal{H}_0\,;\, \mathbf{U}_3}
\end{eqnarray}
where $\mathcal{H}_1$, $\mathcal{H}_2$ and $\mathcal{H}_3$ are tensors of size $n_1\times r_2\times r_3$, $r_1\times n_2\times r_3$ and $r_1\times r_2\times n_3$, respectively.

The second level HMD terms expand upon this concept to capture pairwise interactions. By systematically deflating the core and first level terms, these equations isolate complex joint distributions, such as spatial-spatial correlations or cross-domain spatial-spectral dependencies embedded directly within the HS cube under consideration. Consequently, the second degree HMD components are computed in a similar fashion as
\begin{eqnarray}
\label{eq:Xij}
\mathcal{H}_{1,2} = \frac{1}{n_3} \kruskal{ \mathcal{H}; \mathbf{U}_3} -
\kruskal{ \mathcal{H}_0; \mathbf{U}_1, \mathbf{U}_2} - \kruskal{\mathcal{H}_1; \mathbf{U}_2 } - \kruskal{\mathcal{H}_2; \mathbf{U}_1 } \nonumber \\
\mathcal{H}_{1,3} = \frac{1}{n_2} \kruskal{ \mathcal{H}; \mathbf{U}_2} -
\kruskal{ \mathcal{H}_0; \mathbf{U}_1, \mathbf{U}_3} - \kruskal{\mathcal{H}_1; \mathbf{U}_3 } - \kruskal{\mathcal{H}_3; \mathbf{U}_1 } \nonumber \\
\mathcal{H}_{2,3} = \frac{1}{n_1} \kruskal{ \mathcal{H}; \mathbf{U}_1} -
\kruskal{ \mathcal{H}_0; \mathbf{U}_2, \mathbf{U}_3} - \kruskal{\mathcal{H}_2; \mathbf{U}_3 } - \kruskal{\mathcal{H}_3; \mathbf{U}_2 }
\end{eqnarray}
where $\mathcal{H}_{1,2}$, $\mathcal{H}_{1,3}$ and $\mathcal{H}_{2,3}$ are tensors of size $n_1\times n_2\times r_3$, $n_1\times r_2\times n_3$ and $r_1\times n_2\times n_3$, respectively.

Finally, by defining the index set $S = \{1, 2, 3\}$, the entire HMD expansion can be re-expressed in an elegant and compact formulation, which is %
\begin{equation}
\label{eq:HMDcompact}
\mathcal{H} = \bigsum_{\mathcal{I}\, \subseteq\, S} \kruskal{ \mathcal{H}_\mathcal{I}\,;\, \mathbf{U}_{\mathcal{I}}}; \qquad S = \{1, 2, 3\}
\end{equation}
In practical HS applications, computing the complete expansion is often computationally expensive \cite{tuna_tgrs,enis}. By systematically truncating (\ref{eq:HMDcompact}) and omitting higher level interaction terms as discussed above, a highly efficient representation containing compressed components of interrelational effects is obtained. This guarantees the retention of essential spatial–spectral multivariance features in contrast to conventional tensor decomposition techniques.

\section{Implementations}
\label{sec:results}

\subsection{Experimental Setup}

To rigorously evaluate the classification performance of the proposed HMD framework, comprehensive experiments were conducted across four HS datasets acquired by various sensors. These datasets are Indian Pines, Pavia University, Botswana, and Dioni HS scenes. The specific properties of these datasets are detailed in Table \ref{tab:datasets}.
\begin{table}[h!]
\centering
\begin{tabular}{cccc}
\hline
Dataset & Sensor & Size & Classes \\
\hline
Indian Pines & AVIRIS & $145\times 145\times 200$ & 16 \\
Pavia University & ROSIS & $610\times 340\times 103$ & 9 \\
Botswana & HYPERION & $1476\times 256\times 145$ & 14 \\
Dioni & HYPERION & $250\times 1376\times 176$ & 14 \\
\hline
\end{tabular}
\caption{Specifications of the HS datasets employed in the experimental evaluation.}
\label{tab:datasets}
\end{table}
Initially, the feature extraction capabilities of the zeroth, first, and second level HMD approximants (HMD-0, HMD-1, and HMD-2) were benchmarked against widely used tensor decomposition techniques which are the Tucker and CP models. To assess the frameworks under varying levels of data compression, the subspace dimensionality reduction parameter $r$ was systematically analyzed within the range of $r \in [2, 10]$. In this initial phase, all evaluation metrics were reported using a Support Vector Machine (SVM) classifier. As the second experimental setup, the extracted spatial–spectral features were evaluated using a diverse set of supervised learning algorithms to ensure robust generalization. The selected classifiers included an Adaptive Boosting (AdaBoost) algorithm configured with a maximum of $20$ splits and $30$ learning cycles, a Support Vector Machine (SVM) utilizing a cubic polynomial kernel, a $k$-Nearest Neighbors (KNN) classifier configured with a single neighbor and Euclidean distance, and Linear Discriminant Analysis (LDA). For this second set of experiments, the dimensionality reduction parameter was fixed at $r = 5$ to demonstrate the framework's efficacy at a moderate compression level. The quantitative classification performance was measured using Overall Accuracy (OA), Average Accuracy (AA), F1-score (F1), and the Kappa coefficient ($\kappa$). To guarantee reliable results, each experiment was repeated ten times, with the corresponding average metrics reported. Additionally, a $5$-fold cross-validation procedure was applied uniformly across all experimental trials.

It should be noted that the support matrices $\mathbf{U}_i$'s were formed from the leading right singular vectors of the respective matrix unfoldings of $\mathcal{H}$. Upon determining these singular vectors, each was normalized via multiplication by $\sqrt{n_i}$ to ensure compliance with the condition in (\ref{eq:Sorth}).

\subsection{Results and Comparative Analysis}

To illustrate the efficiency of the proposed HMD method, experiments were initially conducted for rank values ranging from $r = 2$ to $r = 10$. Performance comparisons with the Tucker and CP methods were carried out to better highlight the superior feature extraction capability of the HMD framework.

\begin{figure}[!hb]
\centering
\includegraphics[scale=.45]{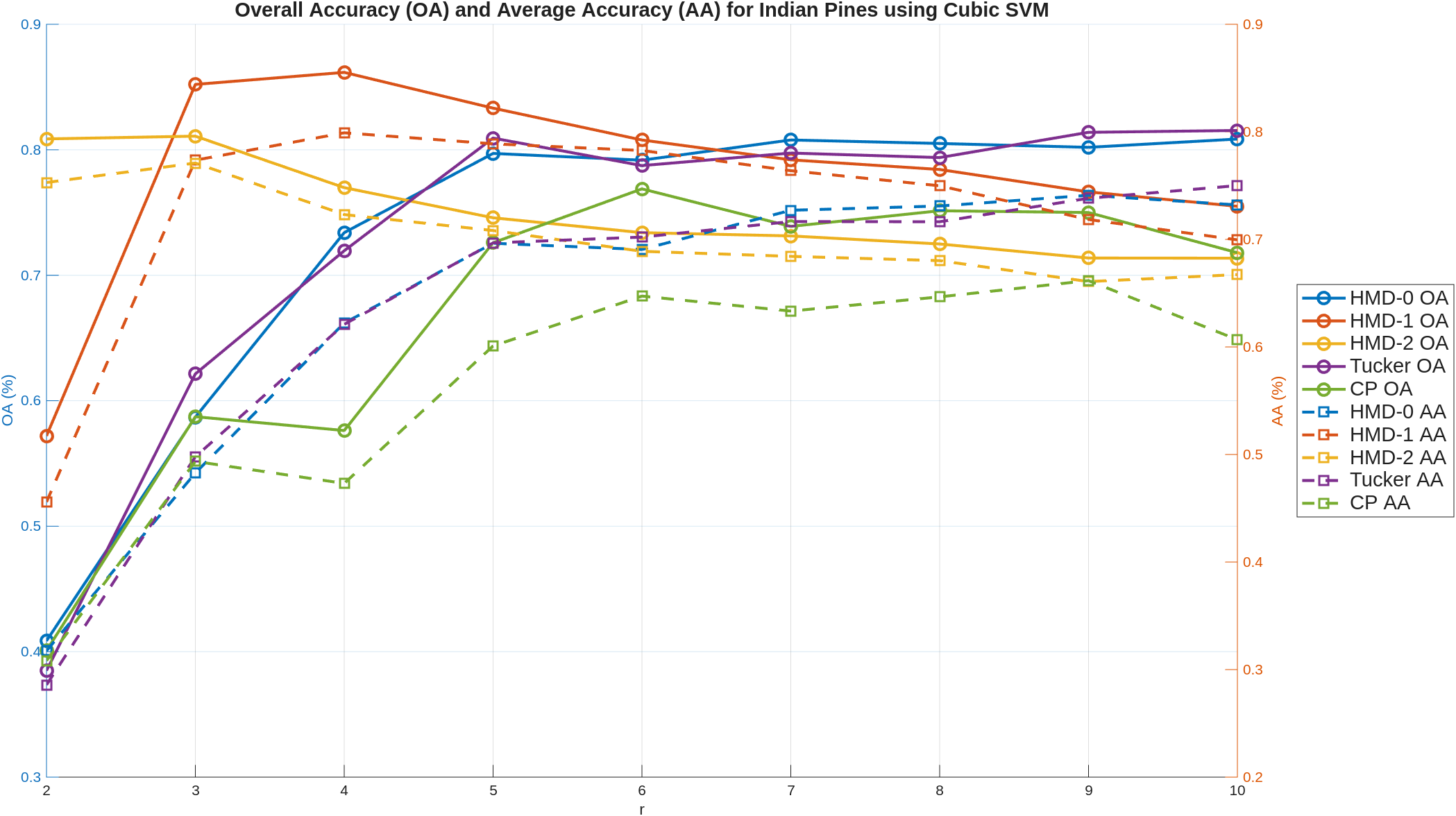}
\caption{OA and AA across different dimensionality reduction parameters ($r$) on the Indian Pines dataset using a Cubic SVM classifier.}
\label{fig:indian}
\end{figure}

Fig. \ref{fig:indian} shows that the first level HMD approximation (HMD-1) achieves superior classification performance, peaking early at $r = 4$. This rapid convergence demonstrates that isolating distinct linear spatial–spectral variations yields a highly discriminative feature set at low ranks. While second level HMD approximant (HMD-2) performs strongly under extreme compression ($r = 2$), its accuracy degrades as the subspace expands, likely due to the introduction of noise from higher order interactions. The zeroth level HMD (HMD-0) and Tucker models exhibit a correlated, gradual ascent that plateaus at higher ranks, whereas the Canonical Polyadic (CP) decomposition consistently underperforms.

\begin{figure}[!hb]
\centering
\includegraphics[scale=.45]{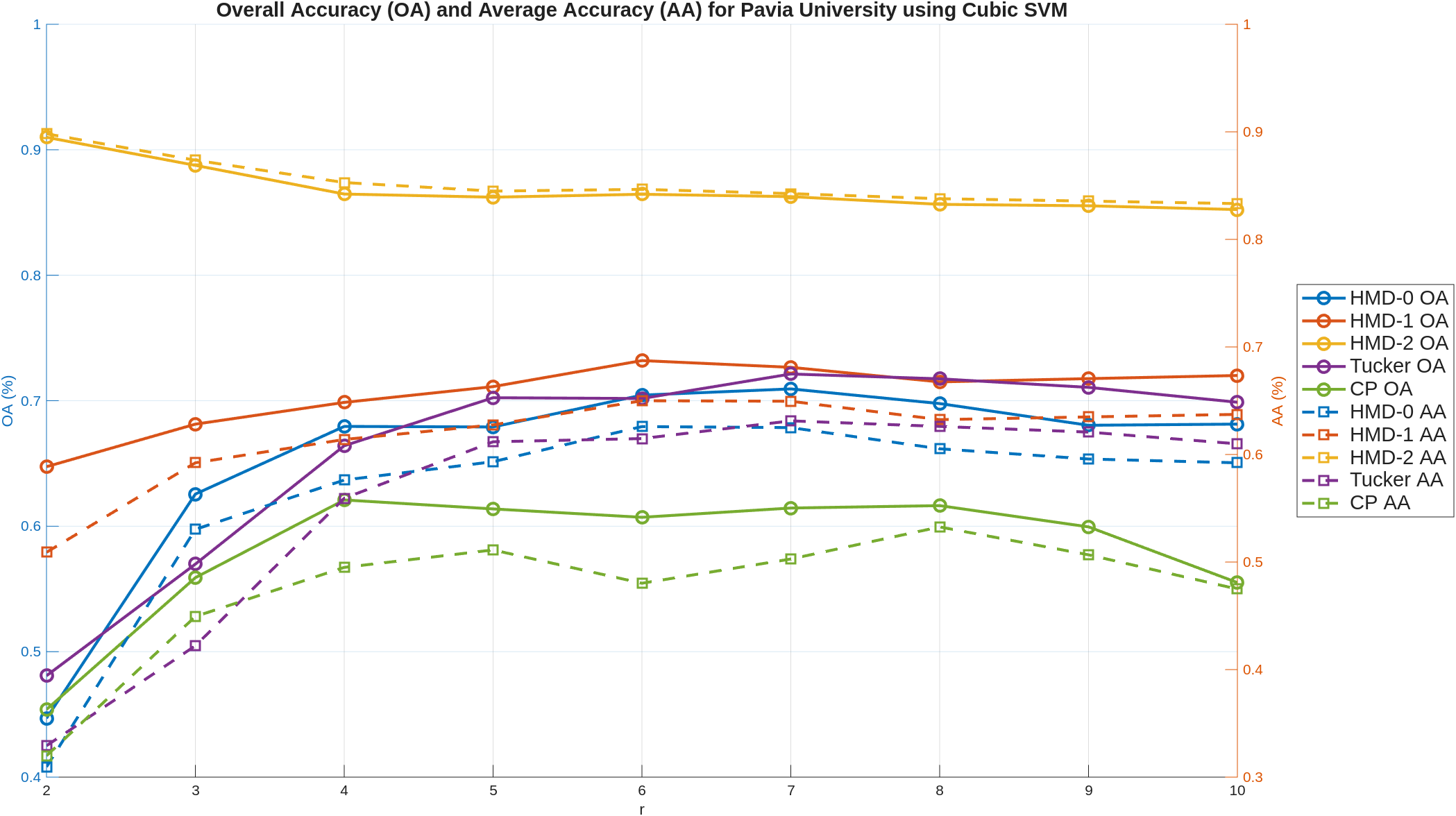}
\caption{OA and AA across different dimensionality reduction parameters ($r$) on the Pavia University dataset using a Cubic SVM classifier.}
\label{fig:pavia}
\end{figure}

According to the observations in Fig. \ref{fig:pavia}, HMD-2 exhibits overwhelming dominance across the entire examined spectrum for the Pavia University scene, peaking at $r = 2$ and maintaining its high accuracy. This underscores that capturing complex pairwise cross-domain interactions is critical for distinguishing highly textured, high resolution urban geometries. HMD-1 provides a stable secondary baseline, while HMD-0 and Tucker require significantly larger subspaces ($r \geq 5$) to converge. CP persistently yields the lowest accuracy, reinforcing its structural limitations.

\begin{figure}[!hb]
\centering
\includegraphics[scale=.45]{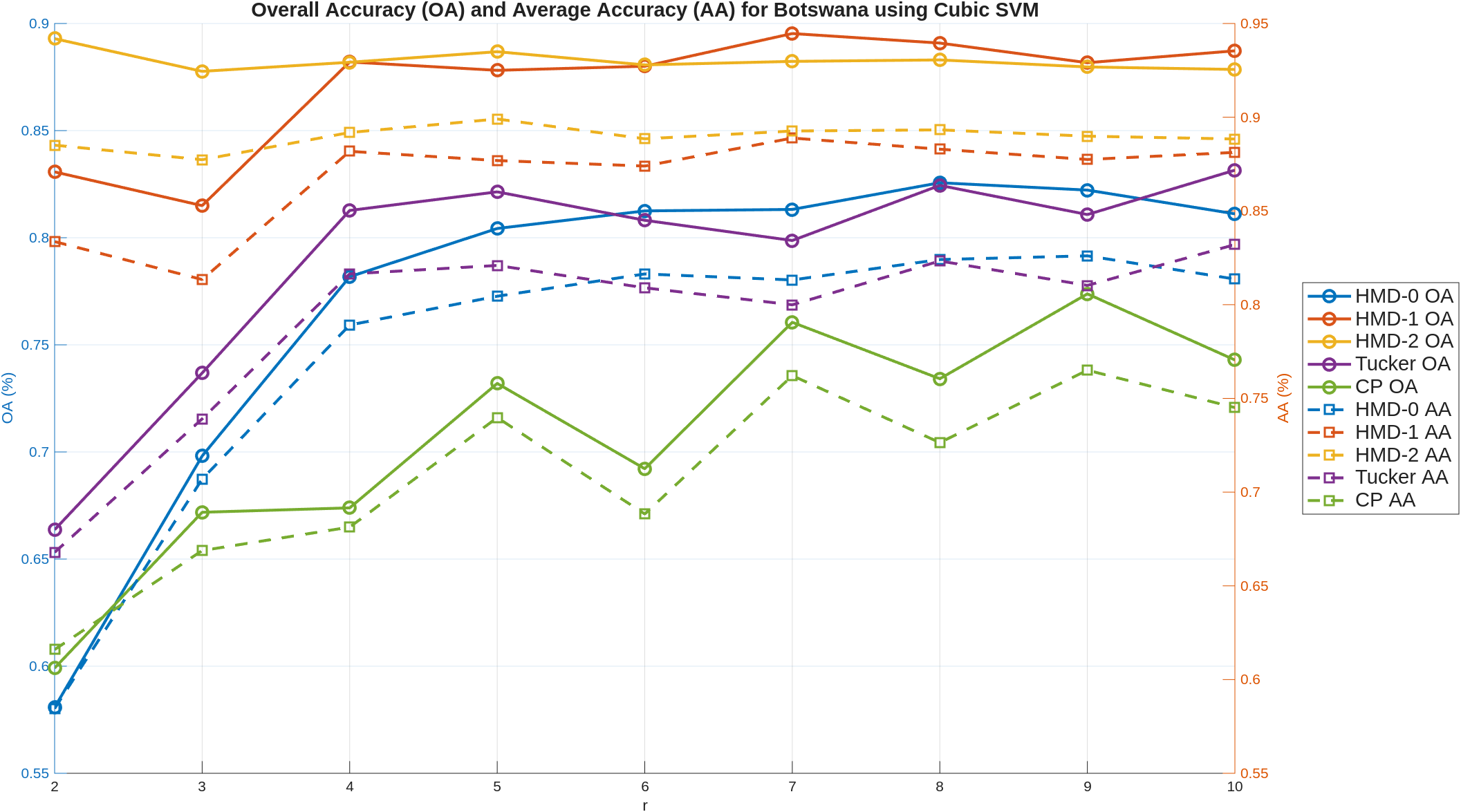}
\caption{OA and AA across different dimensionality reduction parameters ($r$) on the Botswana dataset using a Cubic SVM classifier.}
\label{fig:botswana}
\end{figure}
\begin{figure}[!hb]
\centering
\includegraphics[scale=.45]{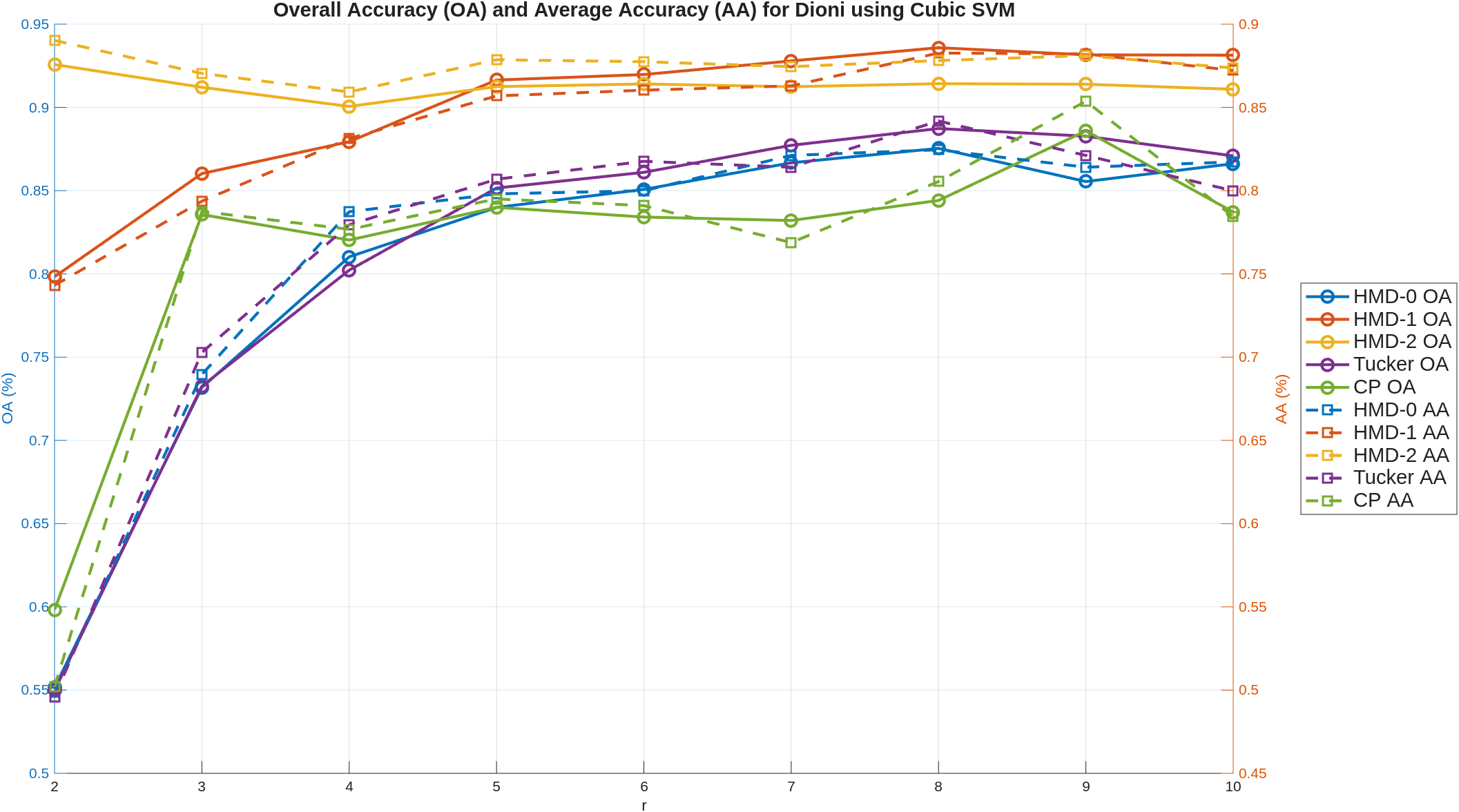}
\caption{OA and AA across different dimensionality reduction parameters ($r$) on the Dioni dataset using a Cubic SVM classifier.}
\label{fig:dioni}
\end{figure}

Both the Botswana and Dioni datasets reveal a dynamic performance crossover between the proposed frameworks as it can be seen in Fig. \ref{fig:botswana} and Fig. \ref{fig:dioni}, respectively. In both scenes, HMD-2 establishes a dominant baseline at the most restricted rank ($r = 2$), effectively leveraging complex joint distributions under severe compression. However, as the allowable subspace expands ($r \geq 4$ for Botswana and $r \geq 5$ for Dioni), HMD-1 rapidly overtakes HMD-2, providing a more robust, less noisy feature representation that achieves the highest overall peak accuracies. Consistent with prior observations, HMD-0 and Tucker exhibit a tightly coupled ascent that plateaus below the advanced HMD variants, while CP displays pronounced volatility and fundamentally inferior performance across the analyzed ranks.

In Tables \ref{tab:indian}, \ref{tab:pavia}, \ref{tab:botswana} and \ref{tab:dioni}, we report the classification metrics obtained using the AdaBoost, SVM, KNN, and LDA algorithms. Each table is partitioned into four sections, detailing the Overall Accuracy (OA), Average Accuracy (AA), F1-score (F1), and Cohen’s Kappa ($\kappa$) metrics. Within every section, the first row presents the results achieved by the zeroth level HMD, while the second and third rows correspond to the first and second level HMD approximants, respectively. Furthermore, the fourth row reports the performance of the Tucker decomposition, and the fifth row displays the values derived from the Canonical Polyadic (CP) decomposition.

\begin{table}[h!]
\centering
\begin{tabular}{ccccc}
\hline
& AdaBoost & SVM & KNN & LDA \\
\hline

\multirow{5}{*}{OA}
& 0.72123 & 0.82717 & 0.79707 & 0.57157 \\
& \textbf{0.75232} & \textbf{0.90169} & \textbf{0.83317} & 0.84818 \\
& 0.69185 & 0.86366 & 0.74579 & 0.\textbf{91771} \\
& 0.71511 & 0.84666 & 0.80930 & 0.60427 \\
& 0.67092 & 0.81132 & 0.72591 & 0.57105 \\
\hline

\multirow{5}{*}{AA}
& 0.54097 & 0.68459 & 0.69641 & 0.45846 \\
& \textbf{0.65092} & \textbf{0.80171} & \textbf{0.78885} & 0.85917 \\
& 0.57917 & 0.78027 & 0.70824 & \textbf{0.89826} \\
& 0.55001 & 0.70210 & 0.69675 & 0.47743 \\
& 0.48704 & 0.66718 & 0.60090 & 0.43282 \\
\hline

\multirow{5}{*}{F1}
& 0.55758 & 0.70334 & 0.70194 & 0.45010 \\
& \textbf{0.68456} & \textbf{0.81936} & \textbf{0.79446} & 0.77623 \\
& 0.60924 & 0.80077 & 0.71552 & \textbf{0.91596} \\
& 0.57051 & 0.72198 & 0.70273 & 0.45924 \\
& 0.50649 & 0.67943 & 0.60350 & 0.42251 \\
\hline

\multirow{5}{*}{$\kappa$}
& 0.67813 & 0.80201 & 0.76842 & 0.50938 \\
& \textbf{0.71323} & \textbf{0.88762} & \textbf{0.80955} & 0.82755 \\
& 0.64053 & 0.84389 & 0.70891 & \textbf{0.90584} \\
& 0.67110 & 0.82411 & 0.78204 & 0.54859 \\
& 0.61858 & 0.78373 & 0.68671 & 0.50774 \\
\hline
\end{tabular}
\caption{Classification evaluation metrics for the Indian Pines dataset across AdaBoost, SVM, KNN, and LDA classifiers.}
\label{tab:indian}
\end{table}

Table \ref{tab:indian} evaluates the decomposition frameworks on the Indian Pines dataset. The first-level HMD (HMD-1) consistently achieves superior performance with the AdaBoost, SVM, and KNN classifiers, indicating that isolating linear spatial and spectral variations yields a robust and highly separable feature space for these algorithms. Conversely, the second-level HMD (HMD-2) achieves the highest metrics under LDA (e.g., OA of $0.91771$, $\kappa$ of $0.90584$), which heavily benefits from the complex joint distributions captured by second-degree components. The zeroth level HMD (HMD-0) and Tucker models exhibit moderate baseline performance, while the Canonical Polyadic (CP) decomposition consistently yields the lowest metrics, reflecting its structural bottleneck.

\begin{table}[h!]
\centering
\begin{tabular}{ccccc}
     \hline
     & AdaBoost & SVM & KNN & LDA \\
     \hline
     \multirow{5}{*}{OA}
     & 0.57342 & 0.73432 & 0.67908 & 0.58757 \\
     & 0.64960 & 0.84153 & 0.71123 & 0.76133 \\
     & \textbf{0.85110} & \textbf{0.95065} & \textbf{0.86220} & \textbf{0.89953} \\
     & 0.61154 & 0.74624 & 0.70237 & 0.59422 \\
     & 0.57409 & 0.69670 & 0.61376 & 0.57882 \\
	 
	 \hline     
	 \multirow{5}{*}{AA}
     & 0.39378 & 0.65386 & 0.59328 & 0.43381 \\
     & 0.48299 & 0.75287 & 0.62756 & 0.66598 \\
     & \textbf{0.79299} & \textbf{0.93428} & \textbf{0.84489} & \textbf{0.88125} \\
     & 0.43014 & 0.64970 & 0.61181 & 0.44396 \\
     & 0.39469 & 0.58101 & 0.51125 & 0.41284 \\
	 
	 \hline
	 \multirow{5}{*}{F1}     
     & 0.39723 & 0.65839 & 0.59472 & 0.43398 \\
     & 0.50521 & 0.77252 & 0.62838 & 0.67061 \\
     & \textbf{0.80779} & \textbf{0.93826} & \textbf{0.84723} & \textbf{0.87842} \\
     & 0.43662 & 0.66316 & 0.61228 & 0.45136 \\
     & 0.39681 & 0.58967 & 0.50898 & 0.42025 \\
	 
	 \hline
	 \multirow{5}{*}{$\kappa$}     
     & 0.38855 & 0.63717 & 0.57201 & 0.41697 \\
     & 0.50769 & 0.78659 & 0.61531 & 0.67580 \\
     & \textbf{0.79729} & \textbf{0.93440} & \textbf{0.81569} & \textbf{0.86713} \\
     & 0.45760 & 0.64943 & 0.60315 & 0.43791 \\
     & 0.40994 & 0.58411 & 0.48767 & 0.41297 \\
     \hline
\end{tabular}
\caption{Classification evaluation metrics for the Pavia University dataset across AdaBoost, SVM, KNN, and LDA classifiers.}
\label{tab:pavia}
\end{table}

For the Pavia University dataset (Table \ref{tab:pavia}), HMD-2 demonstrates uniform dominance across all classifiers and metrics, peaking with an OA of $0.95065$ and $\kappa$ of $0.93440$ using SVM. This systematic superiority underscores that differentiating high resolution urban geometries relies heavily on the complex cross-domain spatial–spectral interactions explicitly isolated by second-degree multivariance components. HMD-1 remains highly competitive but bounded below HMD-2. As observed previously, HMD-0 and Tucker yield comparable, moderate baselines, while CP remains the least effective framework.

\begin{table}[h!]
\centering
\begin{tabular}{ccccc}
     \hline
     & AdaBoost & SVM & KNN & LDA \\
     \hline
     \multirow{5}{*}{OA} 
     & 0.71926 & 0.81704 & 0.80431 & 0.65734 \\
     & 0.83216 & 0.88197 & 0.87814 & 0.92987 \\
     & \textbf{0.85036} & \textbf{0.90434} & \textbf{0.88683} & \textbf{0.96777} \\
     & 0.74629 & 0.83914 & 0.82142 & 0.66233 \\
     & 0.63277 & 0.80472 & 0.73199 & 0.62265 \\
     
     \hline
     \multirow{5}{*}{AA}      
     & 0.71536 & 0.80395 & 0.80452 & 0.66167 \\
     & 0.83190 & 0.87113 & 0.87688 & 0.92195 \\
     & \textbf{0.85358} & \textbf{0.90826} & \textbf{0.89906} & \textbf{0.96344} \\
     & 0.74359 & 0.83165 & 0.82093 & 0.66187 \\
     & 0.63748 & 0.80675 & 0.73976 & 0.63278 \\
	 
	 \hline     
	 \multirow{5}{*}{F1} 
     & 0.72081 & 0.81258 & 0.80542 & 0.66453 \\
     & 0.83203 & 0.87647 & 0.87568 & 0.92594 \\
     & \textbf{0.85301} & \textbf{0.90980} & \textbf{0.89636} & \textbf{0.96529} \\
     & 0.74572 & 0.83470 & 0.81899 & 0.66325 \\
     & 0.63470 & 0.81219 & 0.73747 & 0.63229 \\
	 
	 \hline
	 \multirow{5}{*}{$\kappa$}      
     & 0.69549 & 0.80155 & 0.78789 & 0.62846 \\
     & 0.81818 & 0.87205 & 0.86798 & 0.92399 \\
     & \textbf{0.83785} & \textbf{0.89634} & \textbf{0.87741} & \textbf{0.96508} \\
     & 0.72492 & 0.82560 & 0.80648 & 0.63394 \\
     & 0.60217 & 0.78829 & 0.70968 & 0.59056 \\
     \hline
\end{tabular}
\caption{Classification evaluation metrics for the Botswana dataset across AdaBoost, SVM, KNN, and LDA classifiers.}
\label{tab:botswana}    
\end{table}

Table \ref{tab:botswana} details the classification performance on the Botswana dataset. Consistent with the scene's intricate spatial–spectral characteristics, HMD-2 achieves the highest scores across all supervised models, notably reaching an OA of $0.96777$ and $\kappa$ of $0.96508$ with LDA. While HMD-1 provides a strong secondary baseline, it does not surpass the discriminative capacity of HMD-2. HMD-0 and Tucker maintain their standard core-tensor baseline accuracies, whereas CP persistently registers the lowest evaluation metrics across all configurations.

\begin{table}[h!]
\centering
\begin{tabular}{ccccc}
     \hline
     & AdaBoost & SVM & KNN & LDA \\
     \hline
     \multirow{5}{*}{OA}
     & 0.67330 & 0.75689 & 0.83993 & 0.53521 \\
     & 0.81680 & 0.89634 & 0.91644 & 0.75085 \\
     & \textbf{0.87879} & \textbf{0.93955} & \textbf{0.91249} & \textbf{0.91895} \\
     & 0.68596 & 0.78250 & 0.85156 & 0.57031 \\
     & 0.68489 & 0.76935 & 0.83999 & 0.55167 \\
	 
	 \hline     
	 \multirow{5}{*}{AA}
     & 0.56866 & 0.74563 & 0.79806 & 0.47105 \\
     & 0.71722 & 0.84193 & 0.85690 & 0.76409 \\
     & \textbf{0.80934} & \textbf{0.90655} & \textbf{0.87861} & \textbf{0.91059} \\
     & 0.59025 & 0.75430 & 0.80685 & 0.49373 \\
     & 0.55404 & 0.72710 & 0.79502 & 0.44550 \\

	 \hline     
	 \multirow{5}{*}{F1}
     & 0.60346 & 0.76607 & 0.80593 & 0.46672 \\
     & 0.74445 & 0.85876 & 0.86972 & 0.72740 \\
     & \textbf{0.83839} & \textbf{0.92405} & \textbf{0.89058} & \textbf{0.92071} \\
     & 0.62970 & 0.77446 & 0.81623 & 0.49471 \\
     & 0.59066 & 0.75917 & 0.80461 & 0.46207 \\

	 \hline     
	 \multirow{5}{*}{$\kappa$}	 
     & 0.58444 & 0.69456 & 0.80116 & 0.40194 \\
     & 0.77091 & 0.87107 & 0.89627 & 0.69036 \\
     & \textbf{0.84866} & \textbf{0.92484} & \textbf{0.89135} & \textbf{0.89931} \\
     & 0.59993 & 0.72651 & 0.81541 & 0.44547 \\
     & 0.59619 & 0.70901 & 0.80150 & 0.41264 \\
     \hline
\end{tabular}
\caption{Classification evaluation metrics for the Dioni dataset across AdaBoost, SVM, KNN, and LDA classifiers.}
\label{tab:dioni}
\end{table}     

As shown in Table \ref{tab:dioni} for the Dioni dataset, HMD-2 maintains a dominant discriminative capacity across the majority of configurations, achieving a peak OA of $0.93955$ and $\kappa$ of $0.92484$ with SVM. Although HMD-1 marginally exceeds HMD-2 in isolated instances (such as under the KNN classifier), HMD-2 broadly provides the most consistent feature representation. The baseline dense core-tensor projections (HMD-0 and Tucker) again plateau at lower accuracies, and CP decisively underperforms, reaffirming its structural inadequacy for deeply coupled multidimensional geometries.

\section{Conclusion}
\label{sec:conc}

In this paper, we introduced a novel tensor decomposition method named Holistic Multivariance Decomposition (HMD) for extracting efficient spatial-spectral features of HS images. The proposed framework operates as an adjustable approximation technique similar to the Tucker decomposition, while simultaneously capturing complex cross-mode interrelations in a manner inspired by EMPR. By uniquely addressing both challenges within a unified mathematical structure, HMD achieves exceptional success in extracting highly discriminative spatial–spectral features from HSI.

Structurally, HMD adopts multiple core tensors whose individual parameter counts remain significantly smaller than that of the original tensor, and it utilizes support matrices with a precisely controllable dimensionality reduction parameter $r$. This inherent structural flexibility allows implementers to dynamically tune the retained subspace information, enabling them to achieve an optimal balance between computational complexity and classification performance.

Our comprehensive numerical experiments demonstrate that HMD provides a robust and dependable feature extraction framework for HSI. When evaluated on various benchmark HS datasets, the higher level HMD approximants (specifically the first level and the second level) consistently yielded superior classification accuracy over conventional Tucker and CP models across diverse supervised learning algorithms, even when the allowable subspace was severely constrained.

The selection of the support matrices remains a crucial step for tailoring HMD to specific data characteristics, whether resolving distinct linear variations or highly textured urban geometries. Consequently, any contextually relevant semi-orthogonal matrix set, such as incomplete Fourier or wavelet bases, can be seamlessly integrated into the framework to further enhance the spatial–spectral feature representation. Given its versatile attributes, structural flexibility, and strong quantitative performance, the proposed HMD framework emerges as a highly competitive and efficient tensor decomposition approach for achieving great classification result by extracting high quality features in HSI.

\bibliographystyle{plainnat}
\bibliography{refs}

@article{tuna_tgrs,
  author={Tuna, Süha and Töreyin, Behçet Uğur and Demiralp, Metin and Ren, Jinchang and Zhao, Huimin and Marshall, Stephen},
  journal={IEEE Transactions on Geoscience and Remote Sensing}, 
  title={Iterative Enhanced Multivariance Products Representation for Effective Compression of Hyperspectral Images}, 
  year={2021},
  volume={59},
  number={11},
  pages={9569-9584},
  doi={10.1109/TGRS.2020.3031016}}

@article{tuna_wavelet_hdmr,
  title={An efficient feature extraction approach for hyperspectral images using Wavelet High Dimensional Model Representation},
  author={Tuna, S{\"u}ha and Korkmaz {\"O}zay, Evrim and Tunga, Burcu and G{\"u}rvit, Ercan and Tunga, M Alper},
  journal={International Journal of Remote Sensing},
  volume={43},
  number={19-24},
  pages={6899--6920},
  year={2022},
  publisher={Taylor \& Francis}
}

@article{enis,
  title={A new feature extraction scheme based on support optimization in Enhanced Multivariance Products Representation for Hyperspectral Imagery},
  author={{\c{S}}en, Muhammed Enis and Tuna, S{\"u}ha},
  journal={Journal of the Franklin Institute},
  volume={362},
  number={2},
  pages={107464},
  year={2025},
  publisher={Elsevier}
}

@article{kolda,
  title={Tensor decompositions and applications},
  author={Kolda, Tamara G and Bader, Brett W},
  journal={SIAM review},
  volume={51},
  number={3},
  pages={455--500},
  year={2009},
  publisher={SIAM}
}

@article{sidiropoulos,
  title={Tensor decomposition for signal processing and machine learning},
  author={Sidiropoulos, Nicholas D and De Lathauwer, Lieven and Fu, Xiao and Huang, Kejun and Papalexakis, Evangelos E and Faloutsos, Christos},
  journal={IEEE Transactions on signal processing},
  volume={65},
  number={13},
  pages={3551--3582},
  year={2017},
  publisher={IEEE}
}

@article{tensor_review,
  author={Wang, Minghua and Hong, Danfeng and Han, Zhu and Li, Jiaxin and Yao, Jing and Gao, Lianru and Zhang, Bing and Chanussot, Jocelyn},
  journal={IEEE Geoscience and Remote Sensing Magazine}, 
  title={Tensor Decompositions for Hyperspectral Data Processing in Remote Sensing: A comprehensive review}, 
  year={2023},
  volume={11},
  number={1},
  pages={26-72},
  doi={10.1109/MGRS.2022.3227063}}

@article{tucker_hsi,
  author={Ji, Xiaoxuan and Li, Pengxian and Wang, Jialin and Xu, Shuang and Ji, Teng-Yu and Peng, Jiangjun and Cao, Xiangyong and Meng, Deyu},
  journal={IEEE Transactions on Geoscience and Remote Sensing}, 
  title={Unified Guided Hyperspectral Image Denoising by Continuous Coupled Tucker Decomposition}, 
  year={2026},
  volume={64},
  number={},
  pages={5519813-5519813},
  doi={10.1109/TGRS.2026.3708294}}

@inproceedings{cp_hsi,
  author={Jouni, Mohamad and Mura, Mauro Dalla and Comon, Pierre},
  booktitle={IGARSS 2019 - 2019 IEEE International Geoscience and Remote Sensing Symposium}, 
  title={Hyperspectral Image Classification Using Tensor CP Decomposition}, 
  year={2019},
  volume={},
  number={},
  pages={1164-1167},
  doi={10.1109/IGARSS.2019.8898346}}

@inproceedings{tuna_sparse_hdmr,
  title={Improving sparse coding based hyperspectral image classification via tensor decomposition and oversegmentation},
  author={Tuna, S{\"u}ha},
  booktitle={AIP Conference Proceedings},
  volume={3489},
  number={1},
  pages={280024},
  year={2026},
  organization={AIP Publishing LLC}
}

@inproceedings{efe_fusion,
  title={Enhancing Hyperspectral and Multispectral Image Fusion Using High Dimensional Model Representation},
  author={Kahraman, Efe and Tuna, S{\"u}ha},
  booktitle={2025 9th International Symposium on Innovative Approaches in Smart Technologies (ISAS)},
  pages={1--7},
  year={2025},
  organization={IEEE}
}

@article{hsi_overview,
  title={Overview of hyperspectral imaging remote sensing from satellites},
  author={Qian, Shen-En},
  journal={Advances in hyperspectral image processing techniques},
  pages={41--66},
  year={2022},
  publisher={Wiley Online Library}
}

@article{hs_apps_review,
  title={Hyperspectral imaging and its applications: A review},
  author={Bhargava, Anuja and Sachdeva, Ashish and Sharma, Kulbhushan and Alsharif, Mohammed H and Uthansakul, Peerapong and Uthansakul, Monthippa},
  journal={Heliyon},
  volume={10},
  number={12},
  year={2024},
  publisher={Elsevier}
}

@article{hs_food,
  title={Advanced applications of hyperspectral imaging technology for food quality and safety analysis and assessment: A review—Part I: Fundamentals},
  author={Wu, Di and Sun, Da-Wen},
  journal={Innovative Food Science \& Emerging Technologies},
  volume={19},
  pages={1--14},
  year={2013},
  publisher={Elsevier}
}

@article{artwork,
  title={Hyperspectral imaging combined with data classification techniques as an aid for artwork authentication},
  author={Polak, Adam and Kelman, Timothy and Murray, Paul and Marshall, Stephen and Stothard, David JM and Eastaugh, Nicholas and Eastaugh, Francis},
  journal={Journal of Cultural Heritage},
  volume={26},
  pages={1--11},
  year={2017},
  publisher={Elsevier}
}

@article{drug,
  title={The use of hyperspectral imaging in the VNIR (400--1000 nm) and SWIR range (1000--2500 nm) for detecting counterfeit drugs with identical API composition},
  author={Wilczy{\'n}ski, S{\l}awomir and Koprowski, Robert and Marmion, Mathieu and Duda, Piotr and B{\l}o{\'n}ska-Fajfrowska, Barbara},
  journal={Talanta},
  volume={160},
  pages={1--8},
  year={2016},
  publisher={Elsevier}
}

@article{corneal,
  title={Hyperspectral image enhancement and mixture deep-learning classification of corneal epithelium injuries},
  author={Md Noor, Siti Salwa and Michael, Kaleena and Marshall, Stephen and Ren, Jinchang},
  journal={Sensors},
  volume={17},
  number={11},
  pages={2644},
  year={2017},
  publisher={MDPI}
}

@article{gastric,
  title={Evaluating the identification of the extent of gastric cancer by over-1000 nm near-infrared hyperspectral imaging using surgical specimens},
  author={Mitsui, Tomohiro and Mori, Akino and Takamatsu, Toshihiro and Kadota, Tomohiro and Sato, Konosuke and Fukushima, Ryodai and Okubo, Kyohei and Umezawa, Masakazu and Takemura, Hiroshi and Yokota, Hideo and others},
  journal={Journal of Biomedical Optics},
  volume={28},
  number={8},
  pages={086001--086001},
  year={2023},
  publisher={Society of Photo-Optical Instrumentation Engineers}
}

@article{pattern,
  title={A survey on representation-based classification and detection in hyperspectral remote sensing imagery},
  author={Li, Wei and Du, Qian},
  journal={Pattern Recognition Letters},
  volume={83},
  pages={115--123},
  year={2016},
  publisher={Elsevier}
}

@article{hs_class,
  title={An extensive review of hyperspectral image classification and prediction: techniques and challenges},
  author={Tejasree, Ganji and Agilandeeswari, Loganathan},
  journal={Multimedia Tools and Applications},
  volume={83},
  number={34},
  pages={80941--81038},
  year={2024},
  publisher={Springer}
}

@article{unmixing,
  title={Sparse unmixing of hyperspectral data},
  author={Iordache, Marian-Daniel and Bioucas-Dias, Jos{\'e} M and Plaza, Antonio},
  journal={IEEE Transactions on Geoscience and Remote Sensing},
  volume={49},
  number={6},
  pages={2014--2039},
  year={2011},
  publisher={IEEE}
}

@incollection{pca,
  title={PCA, kernel PCA and dimensionality reduction in hyperspectral images},
  author={Datta, Aloke and Ghosh, Susmita and Ghosh, Ashish},
  booktitle={Advances in Principal Component Analysis: Research and Development},
  pages={19--46},
  year={2017},
  publisher={Springer}
}

@article{lda,
  title={Folded LDA: extending the linear discriminant analysis algorithm for feature extraction and data reduction in hyperspectral remote sensing},
  author={Fabiyi, Samson Damilola and Murray, Paul and Zabalza, Jaime and Ren, Jinchang},
  journal={IEEE Journal of selected topics in applied earth observations and remote sensing},
  volume={14},
  pages={12312--12331},
  year={2021},
  publisher={IEEE}
}

@article{factor,
  title={Dimensional reduction method based on factor analysis model for hyperspectral data},
  author={Li, Na and Zhao, Huijie and Jia, Guorui},
  journal={Journal of Image and Graphics},
  volume={16},
  number={11},
  pages={2030--2035},
  year={2011},
  publisher={Beijing Zhongke Journal Publising Co. Ltd.}
}

@article{dct,
  title={Three-dimensional discrete cosine transform-based feature extraction for hyperspectral image classification},
  author={Prabukumar, Manoharan and Sawant, Shrutika and Samiappan, Sathishkumar and Agilandeeswari, Loganathan},
  journal={Journal of Applied Remote Sensing},
  volume={12},
  number={4},
  pages={046010--046010},
  year={2018},
  publisher={Society of Photo-Optical Instrumentation Engineers}
}

\end{document}